# Plasmonic Janus particles: A perspective on optical manipulation and biomedical applications

*Alemayehu Nana Koya[1,*], Anastasiia Sapunova[2], Nageswar Reddy Sanamreddy[3,9], Yanqiu Zou[4], Qifei Ma[5], Domna G. Kotsifaki[6], Huaizhou Jin[7], Shangzhong Jin[5], Qing Huo Liu[1], Paolo Vavassori[3,10], and Denis Garoli[2,5,8,*]*

[1]*Ningbo Institute of Digital Twin, Eastern Institute of Technology, Ningbo, Zhejiang 315200, P.R. China.*
[2]*Istituto Italiano di Tecnologia, via Morego 30, I-16163 Genova, Italy.*
[3]*CIC nanoGUNE BRTA, Tolosa Hiribidea, 76, E-20018 Donostia-San Sebastian, Spain.*
[4]*College of Optical Science and Engineering, Zhejiang University, Hangzhou 310027, P.R. China.*
[5]*College of Optical and Electronic Technology, China Jiliang University, Hangzhou 3 10018, P.R. China.*
[6]*Photonics Lab, Division of Natural and Applied Sciences, Duke Kunshan University, 8 Duke Ave, Kunshan, Jiangsu Province 215316, P.R. China.*
[7]*Key Laboratory of Quantum Precision Measurement, College of Physics, Zhejiang University of Technology, Hangzhou 310014, P.R. China.*
[8]*Dipartimento di Scienze e Metodi dell'Ingegneria, Università degli Studi di Modena e Reggio Emilia, Via Amendola 2, 42122, Reggio Emilia, Italy.*
[9]*Department of Physics, University of the Basque Country (UPV/EHU), E-20018 Donostia, Basque Country, Spain.*
[10]*IKERBASQUE, Basque Foundation for Science, Plaza Euskadi, 5, E-48009 Bilbao, Spain.*

[*]Email: alemayehu.koya@gmail.com; denis.garoli@unimore.it

## Abstract

The compositional asymmetry of Janus micro- and nanoparticles gives unprecedented opportunities to manipulate such composite particles with different stimuli to achieve enhanced optical, magnetic and photothermal responses, which can be exploited for sensing, phototherapy, and nanoscale robotic applications. This perspective overviews recent advances in optical manipulation of plasmonic Janus particles and their implications for biomedical applications. In particular, a brief summary of optical, plasmonic, and magnetic manipulation of Janus particles of various compositions are presented. Moreover, the potentials of plasmonic and magnetic Janus particles for targeted drug delivery, photothermal therapy, hyperthermia, bio-imaging, bio-detection, and neuromodulation are briefly discussed. Finally, a perspective on the rational design and applications of this particular family of asymmetric particles is forwarded.

**Keywords**: *plasmonic Janus particles*, *optical manipulation*, *drug delivery*, *phothermal therapy*, *hyperthermia*.

## 1 Introduction

Janus particles, named after the two-faced Roman god, have two parts with distinct physical or chemical properties and functionalities [1, 2]. These highly tunable and controllable anisotropic particles offer unique ability to integrate incompatible properties, such as polarity, amphiphilicity, optical, and magnetic properties in a single structure [1]. This customizable capability of Janus particles allows them to exhibit diverse and synergistic behaviors, paving the way for innovative solutions in targeted drug delivery, imaging, and diagnostics [3-6]. As a result, contemporary research on Janus particles is focused on design, synthesis, and engineering of their optical, photothermal, and magnetic properties for biomedical applications [7, 8].

In particular, plasmonic Janus particles have been focus of current research [9, 10] as they use surface plasmon resonance properties to enhance their optical absorption and induce photothermal effects. These pronounced responses can be exploited for controlled injection of functional nanoparticles into living cells [11], heating and stretching of DNA molecules [12], and regulating directional heating [13], which pave the way toward devising novel applications in nanosurgery, sensing, and drug delivery. Although plasmonic Janus particles hold great promise in various applications, they face significant limitations that arise from their manipulation methods and inherent material properties. The existing manipulation techniques often depend on chemical reactions or thermal gradients, which not only reduce precision but are also highly dependent on the intrinsic properties of the carrier fluid [14], where the nanoparticles are often transported to the target site by the carrier fluid, leading to passive delivery that limits both the control and efficiency of the transport process.

Alternatively, given their remarkable biocompatibility, Janus particles based on magnetic materials (such as iron oxide ($Fe_3O_4$)) have shown great promise as energy-to-heat converters under the influence of a magnetic field [15]. As a result, magneto-thermal effects provided by magnetic Janus particles have been an alternative non-invasive delivery strategy, akin to the photothermal effects. Nevertheless, there are plenty of drawbacks to this approach, in particular in nano-therapeutics where high dosages are needed to achieve substantial therapeutic effects [16, 17]. To overcome these limitations, hybrid magneto-plasmonic Janus particles have emerged as a novel concept for multifunctional biomedical applications [15, 18], which features an unparalleled combination of facile functionalization, colloidal stability, optical and magnetic anisotropy, strong ferromagnetic activity, and enhanced optical absorption in the near-infrared region. Thus, the combination of magnetic and optical properties of plasmonic magnetic nanostructures can pave the way for more advanced use of Janus particles.

To take advantage of these intriguing properties of hybrid Janus particles for various applications in biomedicine, it is important to design and synthesize these composite structures with enhanced optical, magnetic, and photothermal responses. To this end, a number of design and synthesis methods have been reported [8, 19, 20]. Furthermore, a plethora of reports on manipulation of Janus particles with various mechanisms, including plasmonic nanotweezers [21], evanescent fields [14], photonic nanojets [22], and magnetic fields [23] imply their potentials for targeted drug delivery and photothermal therapy. As a result of these recent advances in the design principles and manipulation methods of Janus particles, it is important to provide a thorough overview with a particular focus on their implications for biomedical applications.

In this perspective, we present a brief overview of the recent advances in design principles, manipulation techniques and biomedical applications of novel Janus particles (Fig. 1). In this regard, we showcase state of the art of various compositions and configurations of Janus particles with sizes ranging from micro to nanoscale. Moreover, we briefly discuss recent advances in optical, plasmonic, and magnetic manipulation of Janus particles. Furthermore, we outline the potentials of plasmonic Janus particles for biomedical applications and highlight recent works in targeted drug delivery, bio-imaging, bio-detection, photothermal therapy, and neuromodulation. Finally, we forward our viewpoints on the future direction of the field.

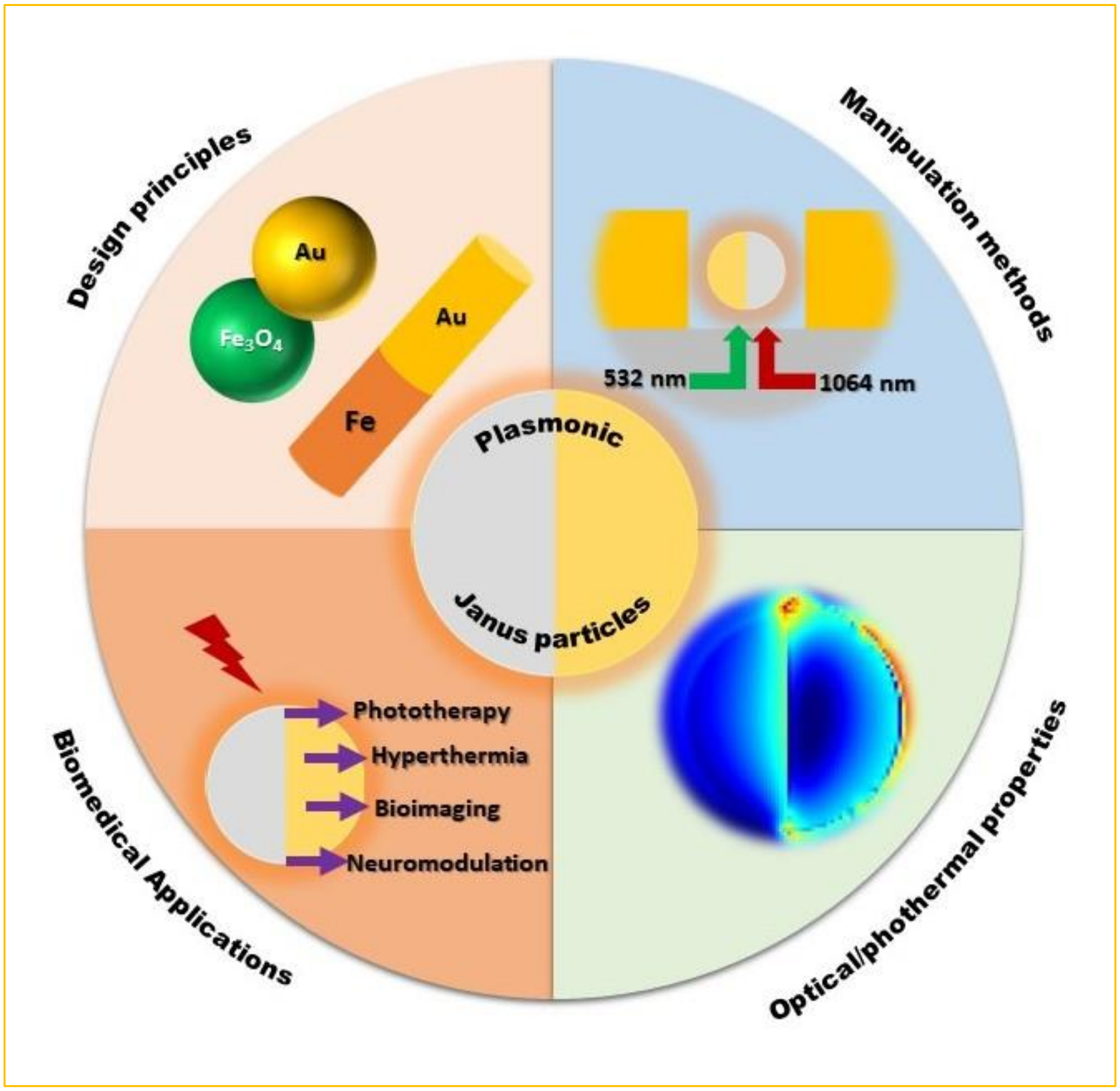

**FIG. 1**. Schematic illustration of the design principles, manipulation methods, optical/photothermal properties, and biomedical applications of Janus particles of various configurations and compositions.

## 2 Optical and magnetic manipulation of Janus particles

The geometrical asymmetry of composite Janus particles gives unprecedented opportunities to manipulate these architectures using various methods including optical manipulation techniques. The interaction of optical fields with Rayleigh particles (particles with sizes much smaller than incident light wavelength) embedded in a medium with the refractive index $n_m$ is determined by its polarizability $\alpha = 3V\frac{\varepsilon_p - n_m^2}{\varepsilon_p + 2n_m^2}$, where $V$ is the particle volume and $\varepsilon_p$ is the dielectric function of the particle [62]. For plasmonic particles, the change in phase of the dipole oscillation with respect to the incident field results in change in sign of the polarizability so that the gradient force is attractive and highly enhanced on the low-frequency side of plasmon resonance but repulsive on the high-frequency side [89]. On the other hand, dielectric particles such as silica beads have positive α, therefore, they are pulled toward the maximum optical intensity of a Gaussian beam with transverse mode [90].

Furthermore, the viscosity of the solvent in which a particle swims plays an important role in determining its acceleration. The limiting velocity $v$ of a particle traveling in a viscous medium can be estimated using Stokes's law: $v = 2qPr/3c\pi\omega_0^2\eta$; where $q$ is the fraction of reflected light, $P$ is the laser power, $r$ is the particle radius, $\omega_0$ is the laser beam radius, and $\eta$ is the viscosity of the medium [91]. The drug force acting on a spherical particle in a viscous fluid moving with speed $v_0$ is given as $F_D = 6\pi\eta r v_0$ [98]. However, optical trapping of plasmonic Janus nanoparticles is challenging, as the high reflectance and absorbance of metals lead to strong forces that repel the particles from the optical trap. To tackle these challenges, recently, several approaches have been devised, including plasmon enhanced resonant optical trapping [21], evanescent field trapping [14], photonic nanojet mediated manipulation [22], and magneto-plasmonic manipulation [23]. Here, we discuss some of these and other related manipulation techniques commonly employed to manipulate plasmonic Janus particles.

### A) Resonant optical trapping of Janus nanoparticles with plasmonic nanoapertures

As a result of the extraordinary optical transmission through nanoholes [24], sub-wavelength nano-apertures in metallic films have been the focus of intense research for the last two decades [25, 26]. In particular, light transmission through circular nanoapertures can be approximated using Bethe's theory [27] and it is inversely proportional to the fourth power of the incident beam wavelength λ, i.e., $T \propto (r/\lambda)^4$, where r is the nanoaperture radius. In presence of an object inside the nanohole, the aperture appears optically larger, which is manifested by redshift in the transmission wavelength and increase in its intensity by ΔT (Figs. 2(a) and (b). Apart from the extraordinary light transmission, plasmonic nanohole apertures (NHAs) are characterized by squeezing photons into a point-like spaces (Fig. 2(c)) [28] and they have remarkable photothermal properties [29]. Because of these intriguing optical and photothermal properties, plasmonic NHAs have found applications in several fields including optical manipulation of small objects with low-power optical tweezers [30-32].

Nevertheless, stable trapping of metallic (and light-absorbing) nanoparticles has been challenging as the high reflectance and absorbance of these materials lead to strong thermal and optical forces that repel the particles from the optical trap [33]. To alleviate the photothermal heating effects of light absorbing nanoparticles, as a principle, an off-resonant optical trapping technique has been widely implemented. But this method ignores self-induced back-action (SIBA) effect, where the trapped objects have an active role in the trapping process [34]. To compensate the drawbacks of off-resonant optical trapping approach, resonant optical trapping method with a two-laser system has been introduced, where one laser excites the trapped objects and the other laser plays the trapping role [35, 36]. In the resonant optical trapping process, the photon energy of the excitation laser matches the excitation bands of the trapped objects, which drastically alters the radiation force through resonantly enhanced induced polarization [37]. As a result, the resonant optical trapping approach is found to be an efficient mechanism to manipulate plasmonic nanoparticles, which enhance the gradient force of an optical trap and increase the strength of trap potential through surface plasmon resonances of metallic particles [38].

One of the most recent works in this regard is the theoretical study by Koya *et al.*, in which they exploited the effect of highly enhanced near-field intensity of plasmonic nanoaperture tweezers and SIBA impact of trapped objects to demonstrate enhanced manipulation of single plasmonic Janus nanoparticles [21]. Systematic displacement of Au-coated Janus nanoparticles toward the nanoaperture and proper orientation of the metallic coating with respect to the aperture results in pronounced near-field intensity and optical force (Fig. 2(d) – (f)). In the abovementioned nanoparticle-nanoaperture configuration, the resonant optical trapping technique can yield about a three-fold optical force, which is ascribed to the excitation of surface plasmon resonance in plasmonic Janus nanoparticles. This resonant optical trapping method can also be employed to manipulate a dimer-like plasmonic Janus particles made of, for example, overlapping $Fe_3O_4$ nanocube and Au nanosphere (Fig. 2(g)). For about 40 nm sized $Fe_3O_4$@Au, the total optical force acting on the nanoparticle can be greatly enhanced by increasing the trapping laser power (Fig. 2(h)). This effort can be further extended by designing multi-component Janus nanoparticles comprised of plasmonic, magnetic, and low-toxic materials like copper sulfide (Fig. 2(i)). Other groups have also demonstrated that the optical resonance between a light-absorbing particle and trapping laser can give rise to about four-fold enhancement of radiation force, which is attributed to the excitation of surface plasmon resonance [36, 38]. Early works by Shoji *et al*. have also demonstrated that combining the resonant optical trapping with plasmonic effects can yield about 5 to 10 times enhanced radiation force [39]. These efforts imply that plasmon-based resonant optical manipulation techniques can not only enhance trapping efficiency but also can give unprecedented opportunities to manipulate various compositions of Janus particles made of dielectric and metallic components [40].

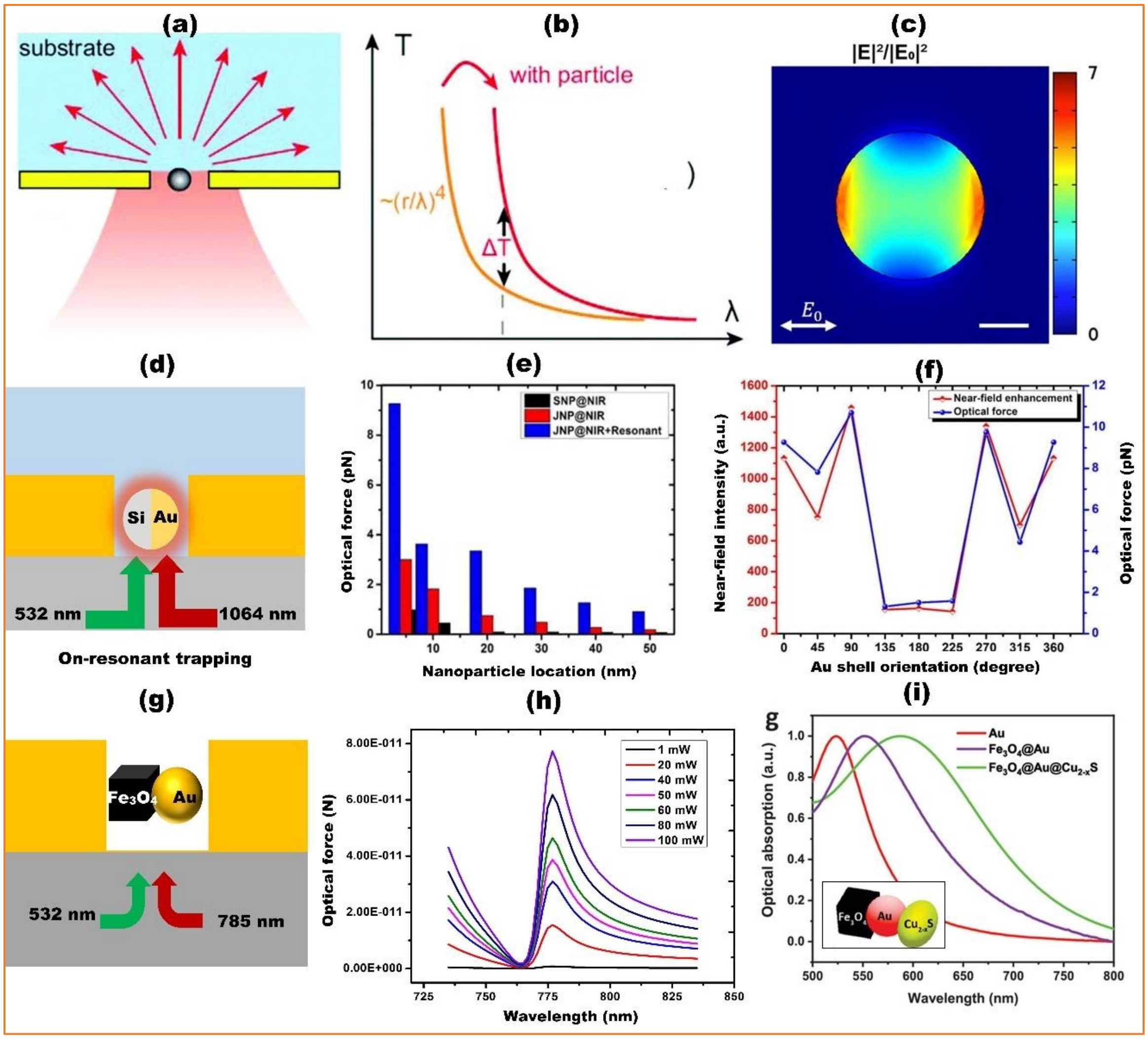

**FIG. 2.** Plasmonic nanohole aperture-enhanced resonant optical trapping of metallic Janus nanoparticles. Illustration of (a) optical transmission through a single nanohole aperture and (b) corresponding transmission intensity with and without particles, where in presence of nanoparticles, the nanoaperture transmission wavelength redshifts and its intensity increases by $\Delta T$. Reproduced with permission from ref [27]. Copyright 2012 Royal Society of Chemistry (c) Typical near-field distribution of single plasmonic nanohole aperture. Reproduced with permission from ref [41]. Copyright 2019 American Chemical Society. (d) Resonant optical trapping of Si-core Au-shell Janus nanoparticle with plasmonic nanoaperture, where Janus nanoparticle location (e) and Au-shell orientation with respect to the plasmonic nanohole aperture (f) greatly affect the optical force strength and near-field intensity. Reproduced with permission from ref [21]. Copyright 2023 AIP Publishing. (g) Resonant optical trapping of $Fe_3O_4$@Au Janus nanoparticle in single nanohole plasmonic aperture with corresponding optical force displayed in (h) for various input laser powers. (i) Optical absorption spectra of hetrostructure Janus nanoparticles made of Au nanoparticles (red line), $Fe_3O_4$@Au dimers (violet line) and $Fe_3O_4$@Au@$Cu_{2-x}$S trimers (green line). Reproduced with permission from ref [78]. Copyright 2022 Wiley-VCH.

## B) Propulsion of Janus particles with optical nanofibers

Optical nanofibers with the waist diameter smaller than the propagating wavelength of light are found to be promising due to their large evanescent field that extends beyond the fiber surface [42]. As a result, optical nanofiber (ONF) tweezers have emerged as powerful optical manipulation tools for trapping of nanoparticles, transportation of living micro-organisms, and propulsion of composite particles [14, 43, 44]. Unlike the free-space beams of light, evanescent fields of optical fiber facilitate stable trapping of Janus particles with low optical power. In this regard, Tkachenko *et al*. have demonstrated stable trapping and efficient propulsion of plasmonic Janus particles (JPs) made of microsphere silica core and half-coated Au nanolayer with optical nanofiber tweezers (Fig. 3(a)). The presence of metallic coating gives rise to an intense back-scattering of light (and hence larger transfer of linear momentum along z) compared to the bare silica particle (SP) (Figs. 3(b) and (c)). The trapping and propulsion actions, respectively, take place via pulling of the JP to the ONF by the radial force $F_y$ and pushing it along z by the longitudinal force $F_z$ (Fig. 3(d)). As a result of plasmonic effect coming from the Au coating, the optical force and propulsion speed of the Janus particle can be significantly enhanced depending on the nanoshell orientation (Figs. 3(e) – (h)). For a given Au nanoshell thickness, the propulsion speed is found to be maximum when the Au cap is oriented toward the nanofiber, which is ascribed to the high polarizability of gold and sharp decline in the evanescent field with distance from the fiber surface [14].

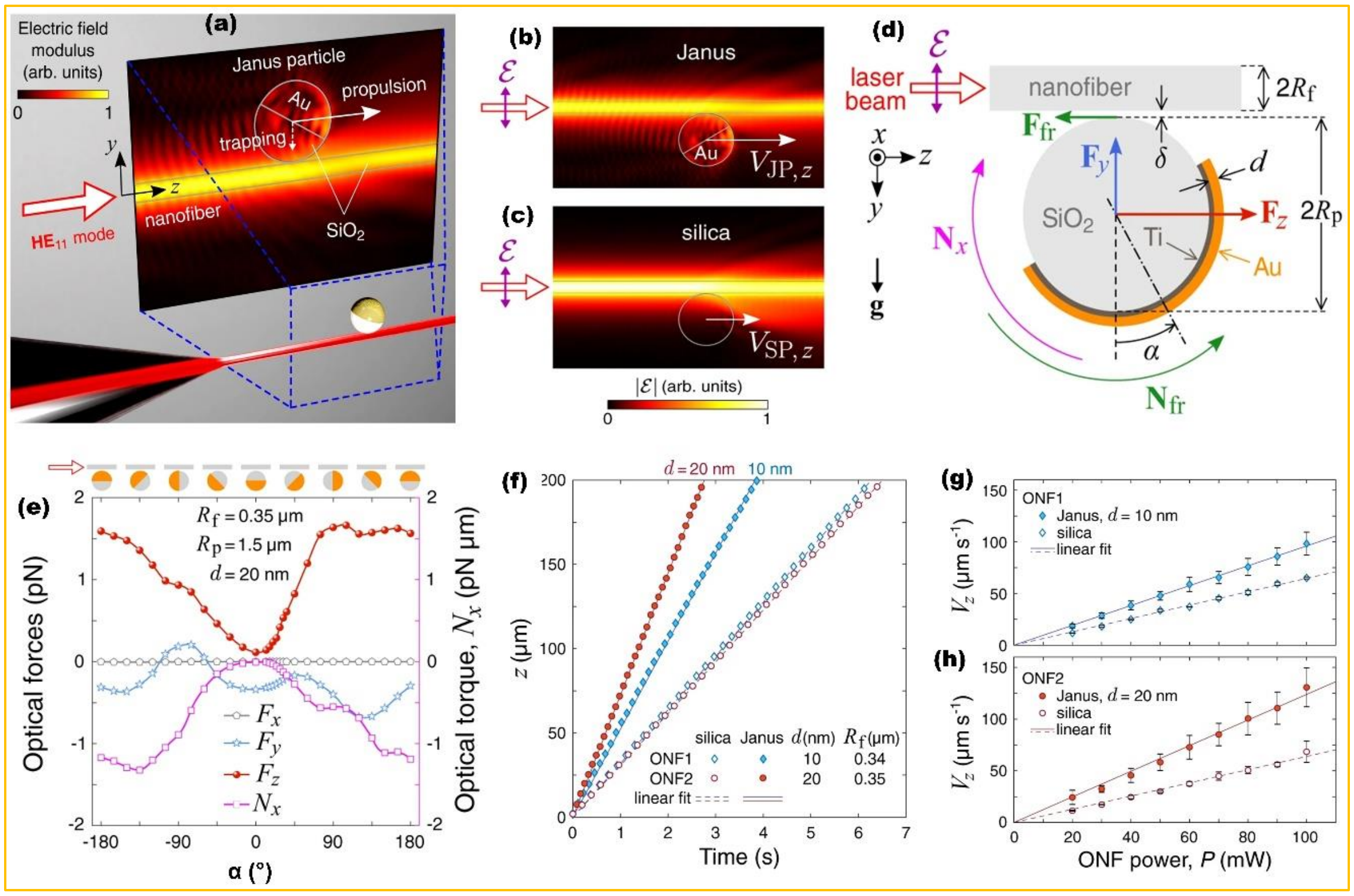

**FIG. 3.** (a) Illustration of trapping and propulsion of plasmonic Janus particles with optical nanofiber (ONF) tweezers. (b) and (c) Comparative electric field distributions around Janus particle (JP) and silica particle (SP) with strong back-scattering of light by the JP. (d) Schematic model of ONF-JP system showing electric field vector ε, gravitational acceleration g; optical torque $N_x$, torque caused by friction force ($F_{fr}$) $N_{fr}$, longitudinal force $F_z$ and radial force $F_y$. (e) The effect of gold cap orientation on the optical force components and optical torque. (f) Position-time series for the longitudinal direction of silica and Janus particles. (g) and (h) Propulsion speed as a function of transmitted optical power for Au coating thickness of d = 10 nm and d = 20 nm, respectively. Reproduced with permission under a Creative Commons Attribution 4.0 International License from ref [14]**.** Copyright 2023 The Authors.

## C) Magnetic manipulation of magneto-plasmonic Janus particles

Magnetic manipulation is another promising technique that allows the precise control of the Janus particles comprised of plasmonic and magnetic components via externally applied magnetic fields. In principle, magneto-plasmonic particles can be designed as a magnetic-core metal-shell bifunctional particles [23] or a Janus-like plasmonic-magnetic particles as illustrated in Fig. 4(a), where the magneto-plasmonic effect is the key property to manipulate with a high degree of accuracy and selectivity [2]. One of the most significant theoretical studies on the possibility of capturing and controlling magneto-plasmonic nanoparticles inside a hybrid magneto-plasmonic nanopore is the recent work by Maccaferri *et al.*, which highlights the prospects of this technique [23]. In this work, the authors theoretically demonstrated that a hybrid nanopore consisting of a thin layer of cobalt sandwiched between two Au layers can be used to create a localized magnetic field by external magnetic field, which allows capturing and control of magnetic nanoparticles near the nanopore (Fig. 4 (b)). According to the simulation results, the capture force of nanoparticles with magnetic tweezers can reach up to 28 pN (Fig. 4(c)), which significantly exceeds the capabilities of traditional optical or plasmon capture methods. There is also an increase in the electromagnetic field in the system (Figs. 4(d) and (e)), which has implication for improved performance in the detection and analysis of molecules. The magneto-plasmonic nanoparticles introduced into these systems can be controlled using an external magnetic field, which allows precise control of their movement through nanopores. This is especially important for applications related to DNA sequencing, where high sensitivity and control over the movement of molecules are required [23]. Even if this model reported only on symmetric magneto-plasmonic particles, a similar effect is expected using a Janus-like magneto-plasmonic particles illustrated in Fig. 4(a). These non-toxic particles can be designed from fluorescent polystyrene nanospheres with a diameter of about 100 nm and several layers of magnetic and plasmonic material coatings on one side of the particle (Fe/Au or Co/Au).

Early works by Erb *et al*. also demonstrated the synthesis of new types of spherical Janus particles that have a metallic coating on less than 20% of the surface (the so-called "dot" Janus particles) [45]. These particles have magnetic anisotropy, which makes them compatible with (magneto-) optical traps. The study showed that such Janus particles can be effectively controlled in three-dimensional space using optical traps, while a magnetic field is used to control the orientation of the particles (Figs. 4 (f) – (h)) [45]. A more recent work by Brenann *et al*. used magneto-plasmonic Janus particles with magnetic core ($Fe_3O_4$) and plasmonic shell (Au) to examine the effect of gold shell thickness on the magnetic properties of the Janus nanoparticles and exploited their potentials for magnetic hyperthermia and photothermal therapy [46]. It was shown that increasing the thickness of the gold shell reduces the magnetic susceptibility of nanoparticles, which negatively affects the ability of nanoparticles to magnetic manipulation and magnetic hyperthermia. With an increase in the thickness of the gold shell, the time of magnetophoresis (*i.e.*, movement of nanoparticles under the influence of a magnetic field) increased significantly, resulting in decrease in the speed and efficiency of magnetic manipulation. This phenomenon is probably due to a decrease in magnetization and an increase in the resistance force during the movement of nanoparticles [46]. Thus, magneto-plasmonic control of composite nanoparticles enables manipulation at the nanoscale, which is important for both the development of new diagnostic and treatment methods and the design of advanced sensors and devices for biological analysis.

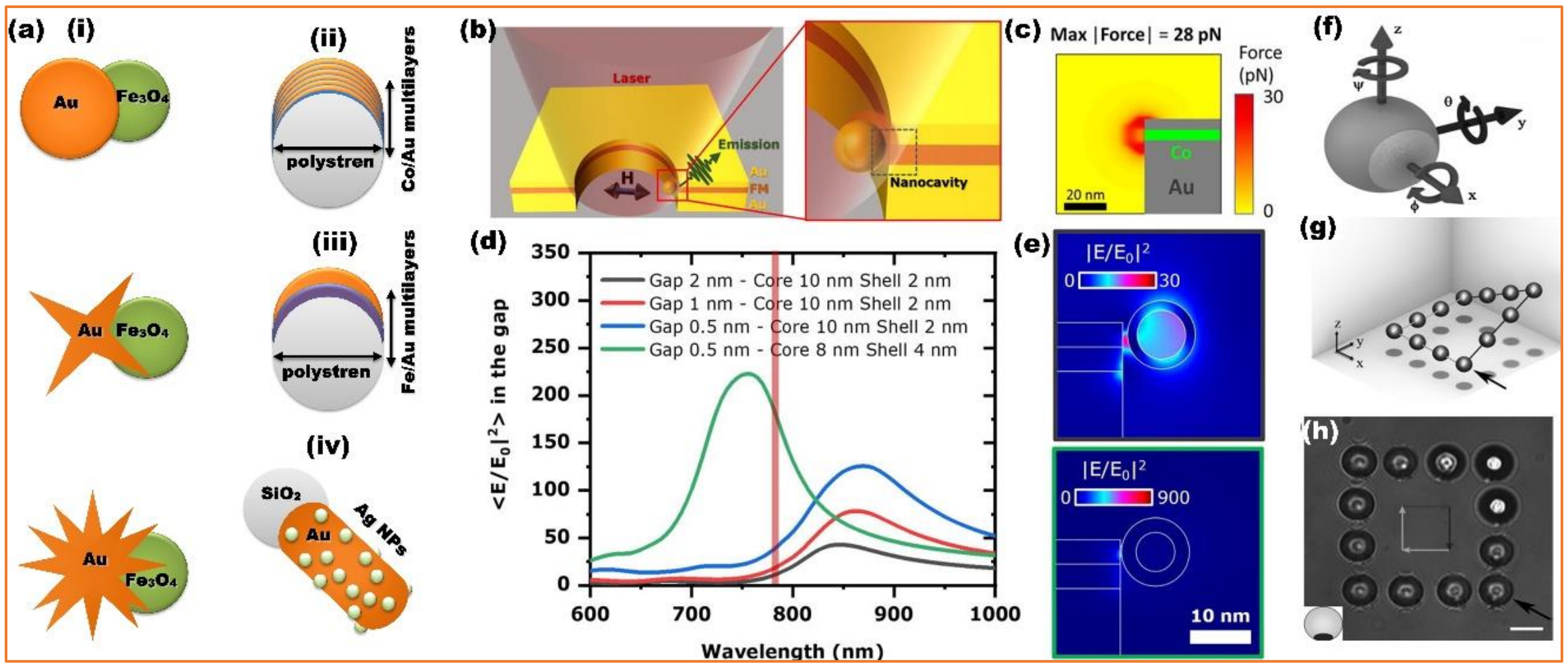

**FIG. 4**. Magnetic control of Janus nanoparticles. (a) A schematic illustration of different designs of plasmonic-magnetic Janus particles: (i) magneto-plasmonic heterodimers comprised of $Fe_3O_4$ and Au in various shapes, (ii) & (iii) magneto-plasmonic nanodomes coated with Co/Au and Fe/Au multilayers, respectively, on polystyrene spheres, (iv) magneto-plasmonic heterodimer composed of $SiO_2$ and Au nanorods that are further coated with Ag nanoparticles. (b) Magneto-plasmonic hybrid nanopore made of Au /Co/Au layers and containing a magnetic core–shell ($Fe_2O_3$-Au) nanoparticle illuminated with a 780 nm laser. (c) 2D map of the force exerted on a 10 nm diameter Magnetite nanosphere. (d) Average field enhancement and (e) near field intensity plots at 780 nm for various nanopore-nanoparticle configurations and geometries. Reproduced with permission under a Creative Commons Attribution (CC BY) License from ref [23]. Copyright 2021 Authors. (f) Dot Janus particle made of polystyrene core and cobalt dot and defined by coordinate axes. (g) Illustration of bead trajectory and (h) micrograph depicting an overlay of 12 images during optical manipulation of the dot Janus particle with magnetic moment. Inset in (h) shows the orientation of the dot Janus particle. Reproduced with permission from ref [45]. Copyright 2009 Wiley-VCH.

## 3 Biomedical applications of plasmonic Janus nanoparticles

Recent studies demonstrated that magneto-plasmonic Janus particles based on heterodimer particles consisting of $Fe_3O_4$ and Au of various shapes exhibited enhanced magnetic-photothermal effects, boosting the potential of these multifunctional materials for biological applications [17]. Moreover, decorating magnetic Janus particles with Ag nanoparticles, has enabled the development of Ag-particle delivery systems, known for their low toxicity and antibacterial properties. Hence, the integration of strong magnetic properties with high bacterial affinity in these system enhances the efficiency of bacterial capture and separation[47]. These and other related findings imply that magneto-plasmonic Janus particles have implications for such applications as drug delivery and phothermal therapy. Here, we discuss recent developments in exploiting the potentials of plasmonic Janus particles for biomendical applications.

### A) Self-propelling plasmonic Janus particles for targeted drug delivery

When the energy of incident light matches the resonant frequency of free electrons in nanostructured plasmonic materials, the nanoparticles exhibit strong localized surface plasmon resonance (LSPR). This phenomenon induces two competing optical processes—strong scattering and absorption of the incident light—which are linked to near-field enhancement and the photothermal effect, respectively[48]. As a result, plasmonic nanostructures have been efficient sources of extremely localized heat sources [49-52], which have been exploited for biomedical applications [53].

The key feature of photo-induced plasmonic heating requires remote activation of the heating by applying laser irradiation to plasmonic nanostructures that are designed to optimally convert light into heat [56]. In this regard, various photothermal materials [57] and architectures [58] have been designed as powerful light-to-heart converters. Specifically, Janus-type composite nano-architectures with asymmetric optical and thermal properties (Fig. 5 (a) – (c)) can serve as light-powered self-propelling universal nanomotors and micro-swimmers [59, 60]. The imbalance in the optical and photothermal properties of Janus nanoparticles have been exploited for controlled guiding of composite nanoparticles with light [60, 61]. In this regard, Peng *et al*. have designed dielectric-core Au-shell Janus micro-swimmers driven by a self-sustained electric field that arises from the asymmetric opthothermal response of the particles [55]. The optically generated temperature gradient along the particle surfaces lead to an enhanced opto-thermoelectrical field that propels the particles (Fig. 5(d)), where the swimming direction is determined by the particle orientation. Furthermore, the propulsion speed and rotational rate of the Au-coated Janus micro-swimmers can be further accelerated by increasing the illumination laser power (Figs. 5 (e) and (f)).

However, when the laser power is further increased, a strong heating effect occurs, which causes thermal damage to the plasmonic Janus particles [55, 62]. As a result, measuring the temperature of optically trapped light-absorbing particles has been a subject of intense research [63, 64]. To overcome the trapping laser power-induced heating damage, an off-resonant trapping techniques have been proposed [33]. However, this approach disregards the active role of the particles in the trapping process. Thus, an alternative approach based on a dual laser system where a defocused visible laser excites the nanoparticle and another focused near-infrared laser traps the nanoparticles is found to be a plausible approach for efficient manipulation of plasmonic Janus nanoparticles [21]**.** These light-powered self-propelling plasmonic Janus particles can serve as bases for developing self-propelling nano-carriers and micro-swimmers for targeted drug delivery. To this end, Park and co-workers have developed beta cyclodextrin (β-CD) conjugated Au-$F_e3O_4$ Janus nanoparticles to exploit the chemo-photothermal effects for cancer ablation [92]. These drug-loaded β-CD-conjugated Janus nanoparticles are capable of eliminating cancer from tumor-bearing nude mice without causing adverse effects (Fig. 5(h)).

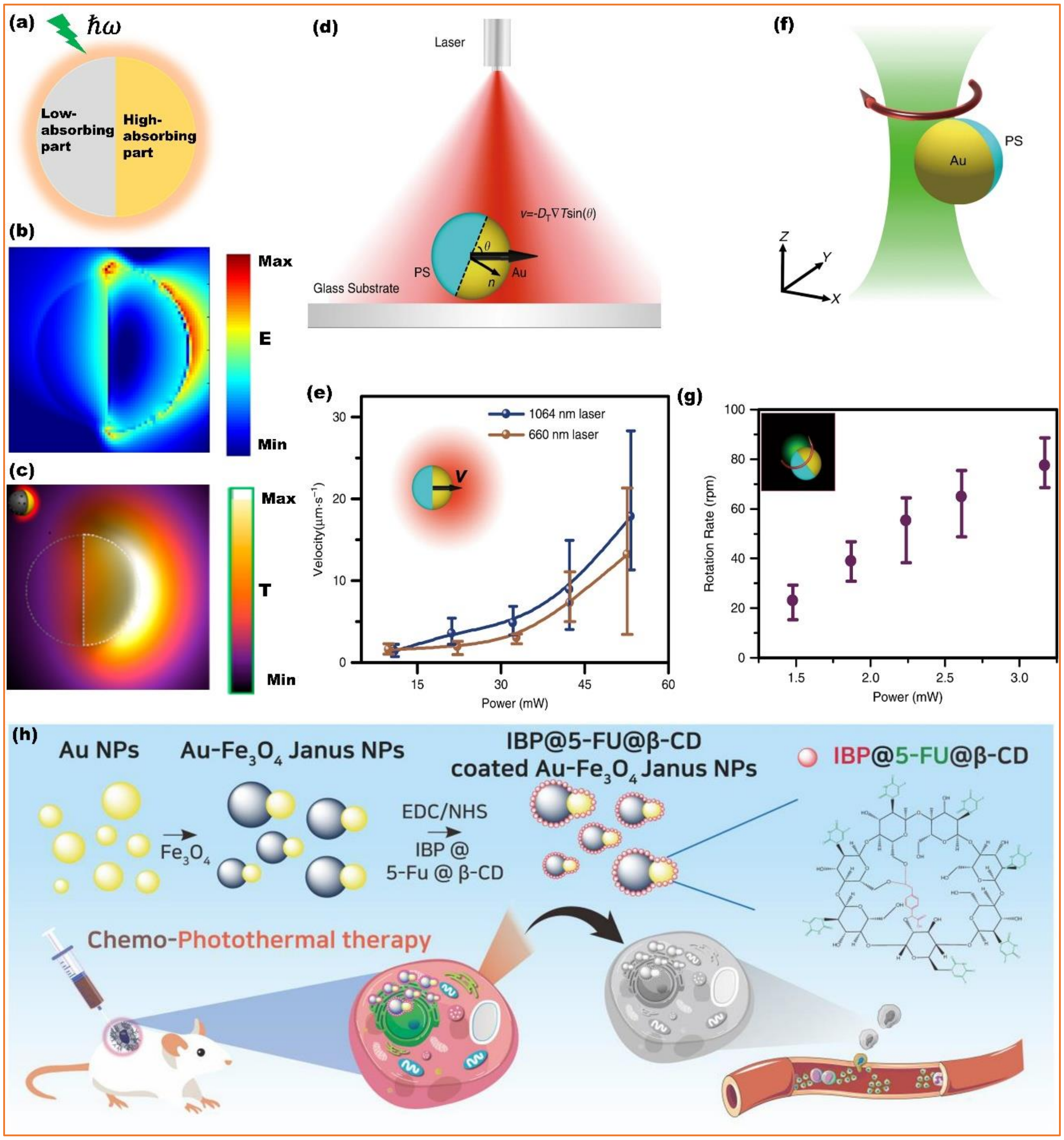

**FIG. 5**. Light-driven self-propulsion of plasmonic Janus particles. (a) Design of plasmonic Janus particles with two different faces made of dielectric core (low-absorbing) and plasmonic shell (high-absorbing) materials. (b) Near-field profile of plasmonic Janus particle. (c) Photothermal asymmetry of plasmonic Janus particles. Reproduced with permission from ref [54]. Copyright 2016 American Chemical Society. (d) Schematic illustration of the light-driven self-propulsion of polystyrene core gold-shell (PS/Au) Janus particle. (e) Swimming velocity as a function of the optical power for 5 µm PS/Au Janus particles for two different laser beams (660 nm and 1064 nm) with a beam size of 31 µm. (f) Schematic illustration of a PS/Au Janus particle rotating under a focused green laser beam. (g) Rotational rate as a function of the optical power for 2.7 µm PS/Au Janus particles. Reproduced with permission under a Creative Commons Attribution 4.0 International License from ref [55]. Copyright 2020 The Authors. (h) Exploiting the potential of drug-loaded plasmonic Janus nanoparticles for cancer eradication from tumor-bearing mice. Reproduced with permission from ref [92]. Copyright 2024 The Authors.

### B) Plasmonic magnetic Janus particles for phototherapy and hyperthermia

#### *Photothermal therapy*

One of the key applications that exploit the photothermal effects of magneto-plasmonic core-shell Janus nanoparticles is cancer therapy and diagnosis [65]. In this regard, Espinosa *et al*. explored the possibility of using magneto-plasmonic Janus nanoparticles combining a magnetic core and a plasmon shell for targeted drug delivery to tumors and subsequent photothermal therapy [66]. The research is devoted to synthesis of heterodimer nanoparticles with an iron core and gold nanostars, thus, the Janus nanoparticles acquire both magnetic and plasmonic properties. After accumulation of the particles in the target area using a magnetic field, the nanoparticles are irradiated with laser radiation in the near infrared range. As a result, the temperature in the particle accumulation zone rises, thereby destroying cancer cells [66]. Brenann *et al*. also used PMJNPs with magnetic core ($Fe_3O_4$) and plasmonic shell (Au) examined the effect of gold shell thickness on the magnetic properties of the Janus nanoparticles and exploited their potentials for photothermal therapy [46].

Huang *et al*. developed $\gamma Fe_2O_3$@Au core/shell-type magnetic gold nanoflower-based theranostic nano-platform, where $\gamma Fe_2O_3$ (magnetic iron oxide) was designed as a core and gold as a shell [67]. These nanoparticles have a nanoflower shape with a rough surface, which enhances surface plasmon resonance and allows them to be used as multifunctional agents for the diagnosis and therapy of cancer. The manipulation of $\gamma Fe_2O_3$@Au particles was carried out by an external magnetic field to accumulate them in the right parts of the body due to the magnetic properties of iron oxide nuclei. The nanoparticles were activated by laser irradiation in the near infrared range, which led to local heating and destruction of tumor cells [67]. Similarly, Liu *et al. also* reported on the development and application of magneto-plasmonic Janus vesicles for improved tumor imaging using magnetic resonance imaging (MRI) and photoacoustic (PA) imaging [68]. The proposed Janus vesicles were created by self-assembling a mixture of magnetic nanoparticles, gold, and amphiphilic block copolymers (BCP). As a drug delivery, the contents of the vesicles were activated by near-infrared (NIR) radiation, and the release rate was controlled by a magnetic field. These vesicles have also proven themselves well as contrast agents for improved imaging of tumors using MRI. The magnetic field significantly increases the accumulation of vesicles in tumors, which led to a 2-3-fold increase in imaging signals compared to the control groups [68].

#### *Hyperthermia*

Hyperthermia is one of the cancer treatment methods that works as an artificial way to increase the temperature of body tissues by heating through external sources, which destroys cancer cells and prevents their further growth [69]. Cancer cells are damaged and destroyed when the temperature reaches around 41 - 42°C, where such cells become vulnerable and can be effectively destroyed, especially by combining other treatment methods like chemotherapy or radiation [70]. However, traditional hyperthermia methods often have limitations, affecting not only the cancerous cells but also the surrounding healthy tissues [71]. With the advent of nanotechnology, particularly Janus nanoparticles, in the process of hyperthermia, it became possible to purposefully control heating, thereby minimizing damage to surrounding healthy tissues. By using Janus particles, several types of hyperthermia can be performed (Fig. 6(a) – (f)). Magnetic hyperthermia involves the use of magnetic nanoparticles that generate heat under the influence of a magnetic field [72]. This method has high selectivity in which magnetic nanoparticles can be purposefully delivered to the tumor using an external magnetic field, which localizes the thermal effect only on unhealthy tissue, minimizing damage to healthy tissue [73]. In related study, Zuo *et al*. used Janus magnetic mesoporous silica nanoparticles (MSNPs) to generate magnetic hyperthermia in combination with protein therapy for enhanced treatment of breast cancer [74]. The dual function of these nanoparticles allowed for both precise heating and targeted delivery of therapeutic proteins, resulting in a significant reduction in tumor growth with combination therapy.

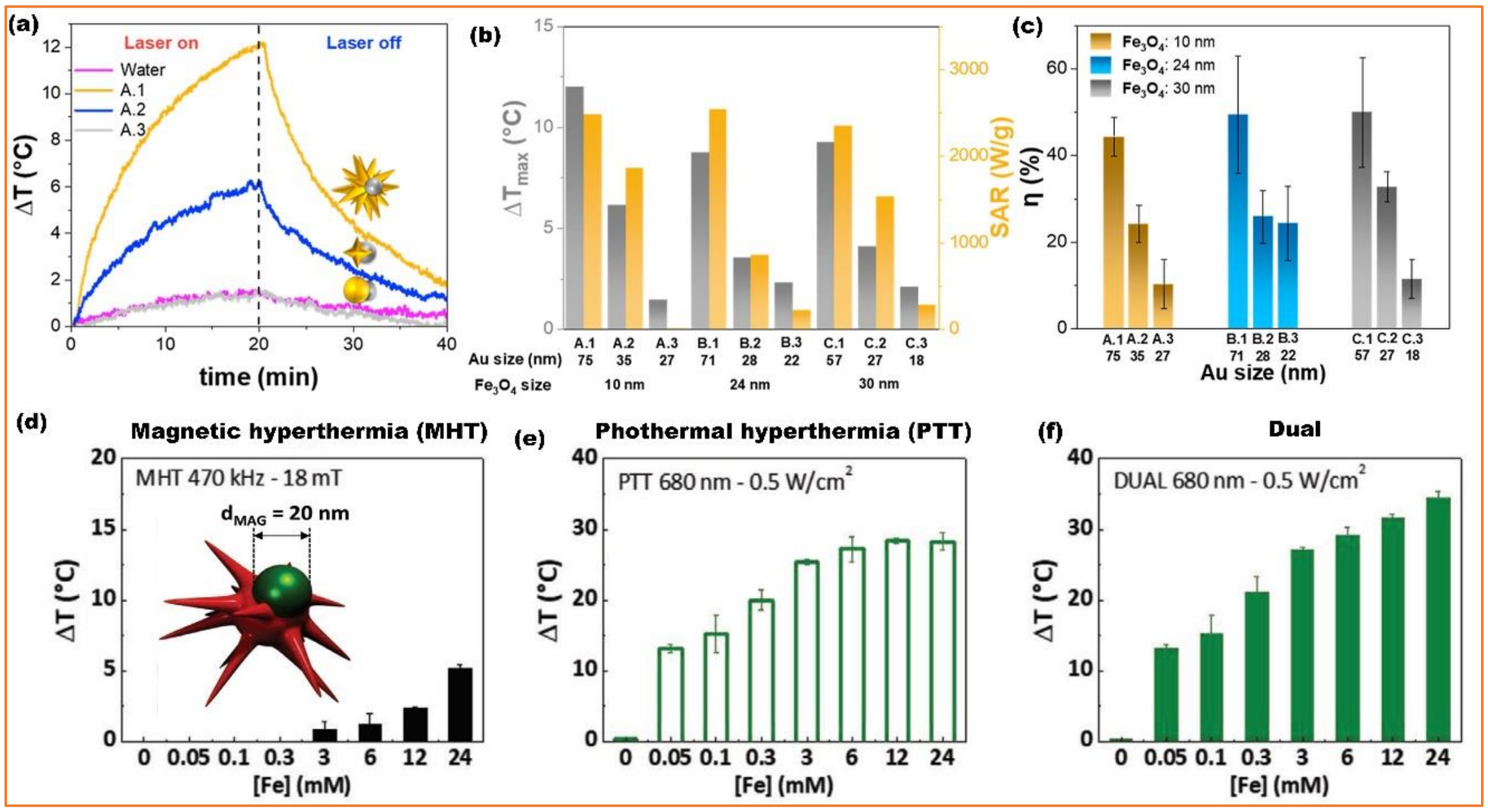

**FIG. 6.** Control of photothermia and magnetothermia through shape and size of various Janus nanoparticles. (a) Thermogram of Au:$Fe_3O_4$ Janus nanoparticles with different sizes and shapes of the Au compositions. (b) Temperature increment and specific absorption rates (SARs) of Au:$Fe_3O_4$ Janus nanoparticles for different sizes of Au and Fe3$O_4$ compositions. (c) Photothermal conversion efficiency (η) of Janus nanoparticles. Reproduced with permission under a CC-BY 4.0 License from ref [17]. Copyright 2023

The Authors. (d – f) Heating properties of Janus magneto-plasmonic nanoparticles made of gold nanostar and iron oxide nanosphere subjected to (d) magnetic hyperthermia (MHT), (e) photothermia (PTT), and (f) both treatments (DUAL) for various iron concentrations. Reproduced with permission from ref [66]. Copyright 2020 Wiley-VCH.

Unlike magnetic hyperthermia discussed above, photothermal hyperthermia uses nanoparticles that absorb light, converting it into heat. While the magnetic hyperthermia is well effective in deeply localized tumors, photothermal hyperthermia is effective for the treatment of superficial tumors or tumors that can be directly irradiated with lasers. The photothermal hyperthermia is enhanced due to the plasmonic properties of Janus nanoparticles. In this regard, Ali *et al.* experimentally investigated the photothermal effects of plasmonic Janus nanoparticles based on silver and nickel (Ag/Ni) and silver and iron (Ag/Fe), which were synthesized using laser ablation [75]. Experiments have shown that the temperature of nanoparticles increases significantly with increasing laser power, where Ag/Ni particles reached temperatures up to 79°C and Ag/Fe up to 72°C at maximum laser power of 4 W. Therefore, exploiting the simultaneous effects of magnetic and photothermal heating increases the hyperthermia of magneto-plasmonic Janus nanoparticles. Such particles could be directed to tumors using magnetic fields, and upon laser irradiation, they generate heat for localized hyperthermia [66]. This dual functionality improves treatment precision and efficacy. In another related report by Wang *et al.* [76], magnetic Janus nanobullets were used to enhance the immune response in cancer therapy by combining magnetic hyperthermia with photodynamic effects. Furthermore, Wu and co-workers also developed Janus nanoparticles with magnetic and luminescent properties for imaging of tumor and the results show decrease in cell viability due to the heat released by the nanoparticles, which makes them applicable for photothermal therapy [77].

Finally, combining hyperthermia with other treatments such as immuno-, chemo-, and radiotherapy provides a multifaceted approach to combating intractable cancers. In this regard, Fiorito *et al*. reported on synthesis of three-material inorganic heterostructures made of iron oxide-gold-copper sulfide ($Fe_3O_4$@Au@$Cu_{2-x}S$) and demonstrated proof of exploitation of the structure for photo-magnetic hyperthermia [78]. The $Fe_3O_4$@Au@$Cu_{2-x}S$ heterostructures with well-defined $Fe_3O_4$, Au, and $Cu_{2-x}S$ domains offer distinct properties for multifunctional cancer treatment. In particular, the $Cu_{2-x}S$ domain contains $^{64}$Cu radioisotope incorporation since it emits both beta particles (for radiotherapy) and positrons (for positron emission tomography (PET) imaging). The study demonstrated that when exposed to both magnetic fields and infrared radiation, the temperature rose higher than usual compared to either of these methods individually. Radiotherapy provides a specific type of treatment because resistant cancer cells may survive hyperthermia, but can be killed by radiotherapy [78]. Thus, the use of magnetic and plasmonic properties of Janus nanoparticles for hyperthermia allows not only to deliver particles to damaged tissues, but also to significantly increase the specificity and effectiveness of hyperthermia.

### C) Plasmonic Janus particles for bio-imaging and bio-detection

On the one hand, visualization of disease tissue with high resolution is of great significance among biomedical applications. Plasmonic-fluorescent Janus particles show great potential for integrating fluorescence and photothermal effect into a single system, implying their potentials for imaging and diagnosis of biological processes [93]. In particular, silver-silver sulfide ($Ag/Ag_2S$) Janus particles have gained a wide-spread attention for bioimaging applications in the second near-infrared (NIR-II) (950−1700 nm) window as a result of the high fluorescence quantum yield and low toxicity of $Ag_2S$. In this regard, Zhang *et al*. developed novel $Ag/Ag_2S$ Janus nanoparticles for activatable fluorescence imaging in the NIR-II window [94]. The excellent fluorescence properties of $Ag_2S$ nanoparticles coupled with the plasmonic properties of Ag can yield hydrogen peroxide ($H_2O_2$) activated NIR-II fluorescence emission of the $Ag/Ag_2S$ Janus nanoparticles (Figs. 7(a) and (b)), which can be employed for *in vivo* NIR-II fluorescence imaging in tumor-bearing living mice (Fig. 7(c)). Similarly, Bao and co-workers designed activatable $Ag_2S$-Ag Janus probes, where the Ag part can yield hydroxyl radicals by consuming $H_2O_2$ for high-efficiency chemodynamic therapy while the $Ag_2S$ part has impressive photothermal therapy (PTT), implying their potentials for precise bio-imaging and synergistic cancer therapy in the NIR-II window [95].

On the other hand, engineering optical hotspots with high density and at the same time concentrating analytes into such hotspots are very important for surface-enhanced Raman scattering (SERS) bio-sensing. This can be realized by using Janus particles as they have two distinct surface regions, which can interact with different biomolecules and membranes. In particular, plasmonic Janus nanoparticles are of interest for detection of small molecules as the plasmonic component generates strong electromagnetic fields that can be exploited for boosting the Raman scattering of surrounding molecules. To this end, Reguera and coworkers demonstrated seed-mediated synthesis of Janus nanoparticle aggregates made of iron oxide nanospheres and gold nanostars (Figs. 7(d) and (e)) to detect dye molecules at low volumes and concentrations [97]. They carried out SERS measurements of magnetically aggregated iron oxide nanosphere-gold nanostar Janus nanoparticles to demonstrate a facile detection of crystal violet [CV] molecules with a concentration of about 450 nM. The magnetic aggregation facilitates concentration of the magneto-plasmonic Janus nanoparticles, yielding ∼5 times enhanced SERS signal (Fig. 7(f)).

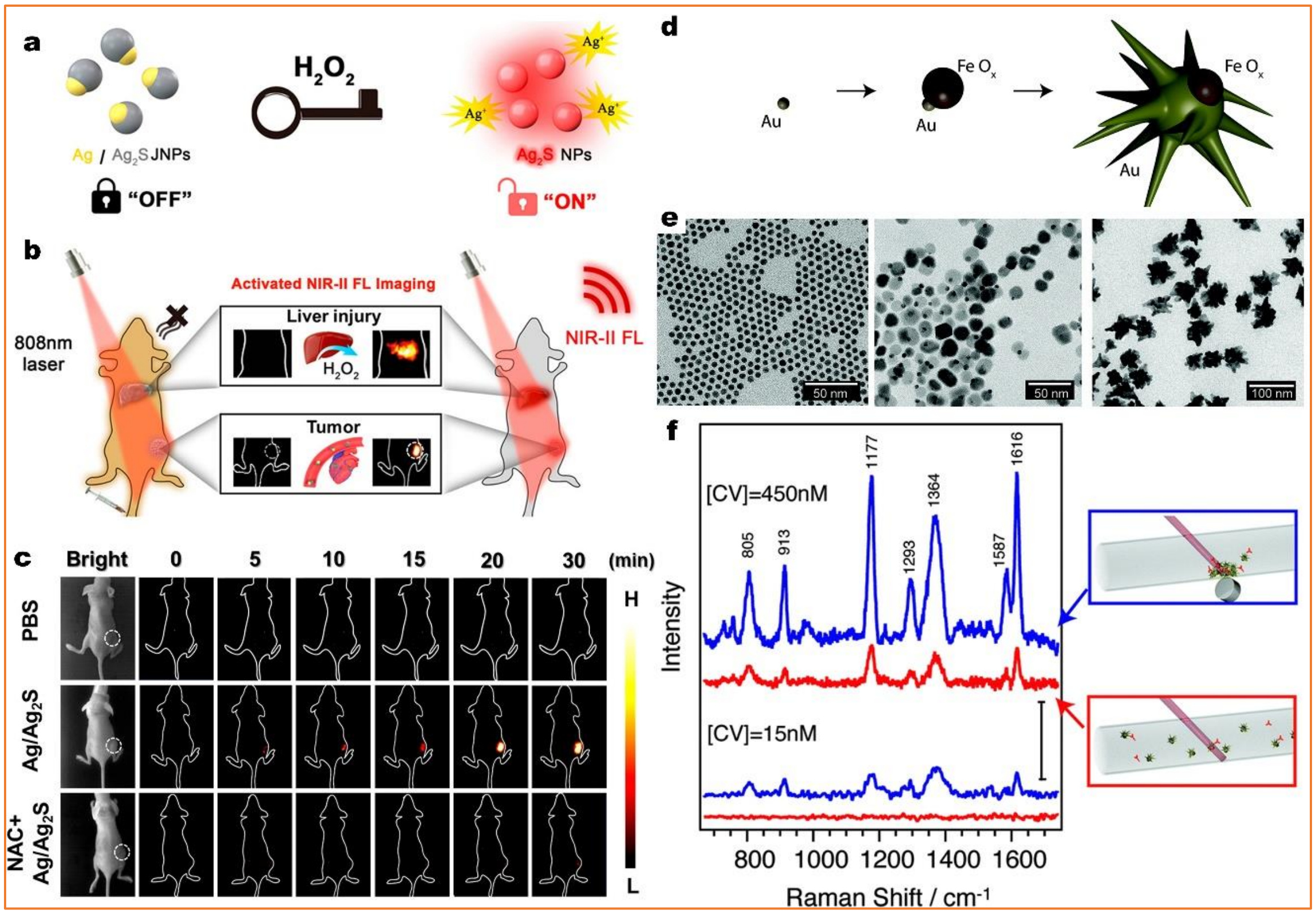

**FIG. 7**. Plasmonic Janus particles for (a - c) bio-imaging and (d-f) bio-sensing applications. (a) Schematic illustration of $H_2O_2$-activated NIR-II fluorescence imaging, where the NIR-II fluorescence signal of Ag/$Ag_2S$ JNPs is switched from "off" to "on" after reaction with the $H_2O_2$. (b) $H_2O_2$-activated Ag/Ag2S Janus nanoparticles for noninvasive detection of the location and diagnosis of acetaminophen-induced liver injury and tumor with high sensitivity and accuracy. (c) $H_2O_2$-activated *in vivo* NIR-II fluorescence imaging in tumor-bearing living mice at 0, 5, 10, 15, 20, and 30 min after Ag/Ag2S JNP injection, where no obvious to small changes occurred in the phosphate buffer (PBS) and N-acetyl-L-cysteine (NAC+Ag/$Ag_2S$) groups. Reproduced with permission from ref [96]. Copyright 2021 American Chemical Society. (d) Scheme of the synthesis of Janus magnetic nanostars through consecutive seed-and-growth-steps. (e) TEM images of different nanoparticles corresponding to the scheme shown in (d). (f) SERS spectra of crystal violet [CV] containing Janus nanoparticles in solution (red) and after magnetic concentration (blue) for two different dye concentrations [CV] = 450 nM (upper spectra) and [CV] = 15 nM (lower spectra). Reproduced with permission under a Creative Commons Attribution 3.0 Unported License from ref [97]. Copyright 2016 The Royal Society of Chemistry.

### D) Magneto-plasmonic Janus particles for neuromodulation

Neural stimulation is an emerging field of study that focuses on treating neurological diseases by exploiting the fundamental wave properties and by engineering nanostructures for efficient modulation of nervous system [79]. Among a wide range of smart nanostructures developed for modulation of neural activities [80, 81], Janus particles provide unique optoelectronic features. To this end, Han *et al.* demonstrated the use of silica-based piezoelectric magnetic Janus microparticles (PEMPs) as miniaturized electrodes to study neural circuits [82]. While the barium titanate ($BaTiO_3$) nanoparticle conjugated half of the PEMP acts as piezoelectric electrode to induce electrical stimulation, the nickel-gold nanofilm-coated magnetic hemisphere provides spatial and orientational control of neural stimulation via external uniform rotating magnetic fields (Figs. 8(a) – (c)). This implies that the anisotropic nature of Janus particles provides unparalleled opportunities for minimal invasive treatment of neurological diseases.

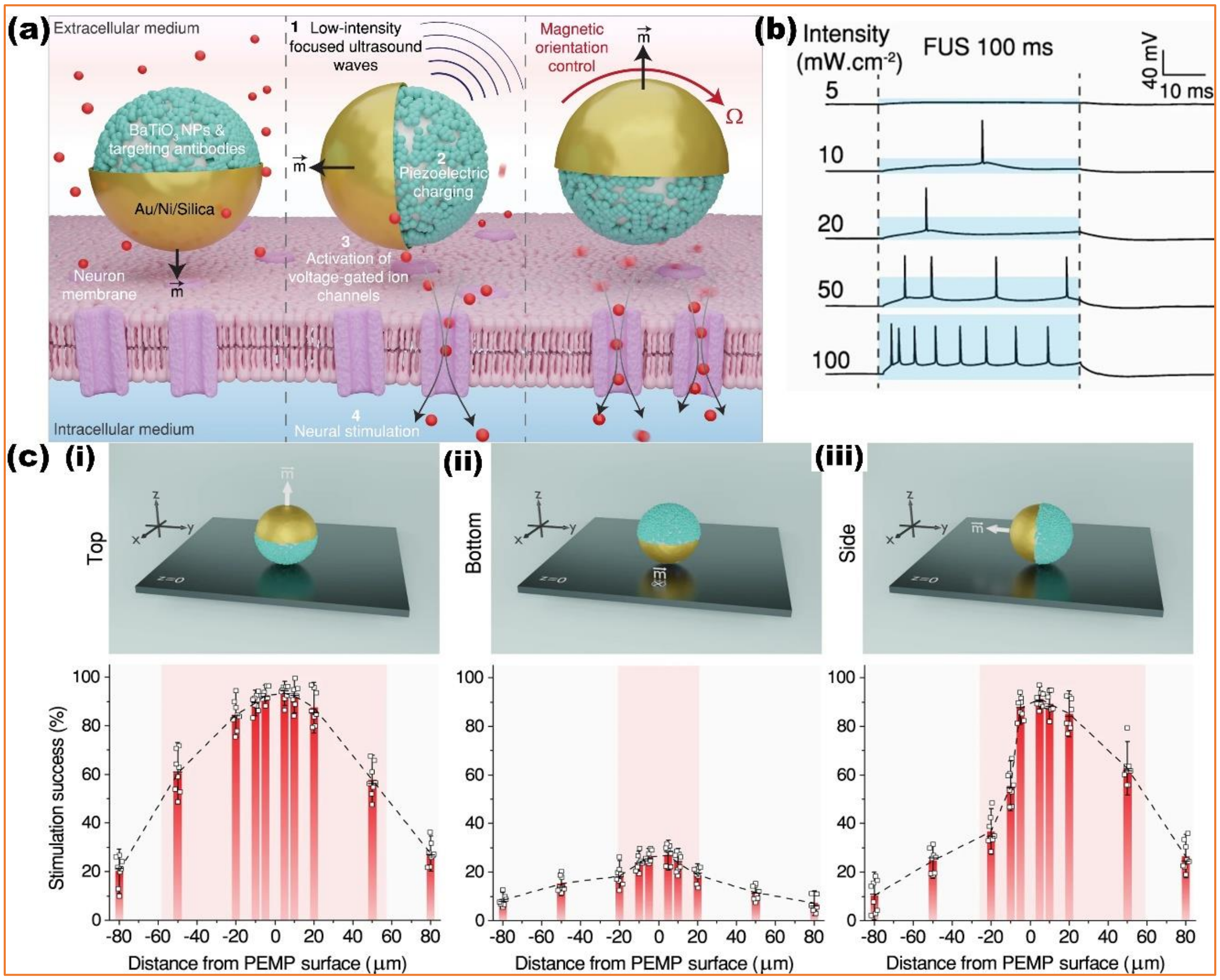

**FIG. 8.** (a) Neuromodulation strategy based on piezoelectric magnetic Janus microparticles (PEMPs). (b) Representative membrane potential traces of primary neurons excited with PEMP under 100 ms focused ultrasound (FUS) with increasing intensities. (c) Top: orientation of a single PEMP is controlled by the out-of-plane rotating magnetic field of 10 mT magnitude with orientation directions depicted as top, bottom, and side; bottom: corresponding neural stimulation success of a single PEMP for neurons calculated at various distances, where the pink box in each plot represents the reproducible high-success stimulation area. Reproduced with permission under a Creative Commons Attribution 4.0 International License from ref [82]. Copyright 2024 The Authors.

Moreover, Mousavi *et al.* provided an in-depth review on the emerging field of neuroplasmonics, emphasizing the critical role of neural interfaces in advancing our understanding of brain function and enhancing neural device technologies [83]. Since, both optical neural interfaces and electrical-based neural interfaces involve intrinsic limitations and challenges, the exploitation of magneto-plasmon Janus-based electrodes holds the potential to transform the brain-machine interface technology. This hybrid strategy, which combines non-genetic magnetic and plasmonic electrical stimulation, has the potential to revolutionize neural engineering [84]. Moreover, these particles could be functionalized with materials created through genetic engineering, providing a novel hybrid (magneto-opto-electro-genetic) neural interface technology. Apart from light and magnetic fields, other forms of physical signals with significant tissue penetration (such as ultrasound) have emerged to enrich the toolbox for ion channel manipulation by acoustic forces that are generated by ultrasound to directly activate mechanosensitive channels with high conductance and Piezo1 (*i.e.*, a mechanosensitive ion channel protein) to excite neurons [85].

## 4 Perspective

In this work, we have presented an overview on the most recent developments in design principles, manipulation methods, and biomedical applications of plasmonic Janus particles. A particular focus is given to novel design of magneto-plasmonic Janus particles with enhanced optical, magnetic, and photothermal responses. Furthermore, we have also overviewed recent advances in optical and magnetic manipulation of Janus particles. We have also briefly discussed the application of these asymmetric particles in the biomedical field, including drug delivery, phothermal therapy, hyperthermia, bio-imaging, bio-sensing, and neuromodulation.

The research interest in exploiting the potentials of Janus particles for biomedical applications is still increasing, thanks to their interesting performance compared to standard symmetric nanoparticles. Compared with isotropic metal or polymer nanoparticles, Janus particles possess two or more domains with anisotropic compositions and independent functions. Although individual components combine and coexist together in a single-particle system, the intrinsic physical and chemical properties of each domain are seldom altered or lost. For instance, Janus particles enable to carry simultaneously multiple drugs and biomolecules. If different domains are modified with light- or heat-sensitive molecules, their release can be separately controlled. Thus, Janus particles with optical and magnetic properties as contrast agents can be used for multimodal bio-imaging, by combining with drug loading and realizing imaging-guided therapy. Finally, Janus particles as biosensor can simultaneously detect many biomolecules, as well as can be used for the construction of self-propelling micro-/nano-motors, achieving targeted drug delivery and cancer therapy.

Despite the important applications of Janus particles in multiple fields from biomedicine to optics and catalysis, there are still considerable room for improvements not only in theoretical understanding of their motion but also in the synthesis methods, propulsion mechanisms, and application aspects. In particular, accurate modeling of the kinetic behaviors of Janus particles in viscous media would shade light on a thorough understanding of the self-propulsion, collective dynamics, and controlled assembly of such particles. However, the physics behind the working mechanism of Janus particle movements and their interaction with environment (typically biologically relevant liquids) is yet to be uncovered. Thus, taking advantage of the cutting-edge simulation tools [14, 86, 87] and state of the art of fabrication methods will provide a comprehensive understanding about the optical, magnetic and thermal properties of Janus particles. Furthermore, having a precise control over the movement and direction of Janus can facilitate their application in targeted drug delivery and nano-robotics. From the applications point of view, one of the main challenges is related to the biodegradability and toxicity of the Janus particles. Thus, developing Janus particles with biodegradable and low-toxic materials that can dissolve harmlessly into their surroundings after completing any desired purpose would benefit the field of biomedicine. Moreover, the use of porous structures can improve the drug loading capacity [88].

Finally, with their asymmetric compositions and distinct chemical and/or physical properties, plasmonic Janus particles are expected to revolutionize not only the field of biomedicine but also its practices. In particular, integrating such smart devices with artificial intelligence will accelerate developing advanced sensing, imaging, and diagnostic tools with enhanced sensitivity, accuracy, and automation. Therefore, we hope that this perspective will stimulate further research efforts to use Janus particles in multiple fields including bioengineering, robotics, and environmental monitoring.

## Acknowledgements

A.N.K. acknowledges support from the Chinese Academy of sciences President's International Fellowship Initiative (Grant No. 2023VMC0020). D. G. thanks the European Union under the Horizon 2020 Program, HORIZON-MSCA-DN-2022: DYNAMO, grant Agreement 101072818.

## References

1. Zhang J, Grzybowski BA, Granick S. Janus Particle Synthesis, Assembly, and Application. Langmuir. 2017;33(28):6964-77.
2. Zhang X, Fu Q, Duan H, Song J, Yang H. Janus Nanoparticles: From Fabrication to (Bio)Applications. ACS Nano. 2021;15(4):6147-91.
3. Tran LT, Lesieur S, Faivre V. Janus nanoparticles: materials, preparation and recent advances in drug delivery. Expert Opin Drug Deliv. 2014;11(7):1061-74.
4. Yi Y, Sanchez L, Gao Y, Yu Y. Janus particles for biological imaging and sensing. Analyst. 2016;141(12):3526-39.
5. Conniot J, Talebian S, Simoes S, Ferreira L, Conde J. Revisiting gene delivery to the brain: silencing and editing. Biomater Sci. 2021;9(4):1065-87.
6. Agrawal G, Agrawal R. Janus Nanoparticles: Recent Advances in Their Interfacial and Biomedical Applications. ACS Applied Nano Materials. 2019;2(4):1738-57.

7. Walther A, Muller AH. Janus particles: synthesis, self-assembly, physical properties, and applications. Chem Rev. 2013;113(7):5194-261.
8. Kirillova A, Marschelke C, Synytska A. Hybrid Janus Particles: Challenges and Opportunities for the Design of Active Functional Interfaces and Surfaces. ACS Applied Materials & Interfaces. 2019;11(10):9643-71.
9. Nedev S, Carretero-Palacios S, Kuhler P, Lohmuller T, Urban AS, Anderson LJ, et al. An Optically Controlled Microscale Elevator Using Plasmonic Janus Particles. ACS Photonics. 2015;2(4):491-6.
10. Bronte Ciriza D, Callegari A, Donato MG, Cicek B, Magazzu A, Kasianiuk I, et al. Optically Driven Janus Microengine with Full Orbital Motion Control. ACS Photonics. 2023;10(9):3223-32.
11. Maier CM, Huergo MA, Milosevic S, Pernpeintner C, Li M, Singh DP, et al. Optical and Thermophoretic Control of Janus Nanopen Injection into Living Cells. Nano Lett. 2018;18(12):7935-41.
12. Simoncelli S, Johnson S, Kriegel F, Lipfert J, Feldmann J. Stretching and Heating Single DNA Molecules with Optically Trapped Gold–Silica Janus Particles. ACS Photonics. 2017;4(11):2843-51.
13. Xie C, Wilson BA, Qin Z. Regulating nanoscale directional heat transfer with Janus nanoparticles. Nanoscale Adv. 2024;6(12):3082-92.
14. Tkachenko G, Truong VG, Esporlas CL, Sanskriti I, Nic Chormaic S. Evanescent field trapping and propulsion of Janus particles along optical nanofibers. Nat Commun. 2023;14(1):1691.
15. Li Z, Lopez-Ortega A, Aranda-Ramos A, Tajada JL, Sort J, Nogues C, et al. Simultaneous Local Heating/Thermometry Based on Plasmonic Magnetochromic Nanoheaters. Small. 2018;14(24):e1800868.
16. Rincon-Iglesias M, Rodrigo I, L BB, Serea ESA, Plazaola F, Lanceros-Mendez S, et al. Core-Shell Fe(3)O(4)@Au Nanorod-Loaded Gels for Tunable and Anisotropic Magneto- and Photothermia. ACS Appl Mater Interfaces. 2022;14(5):7130-40.
17. Abu Serea ES, Orue I, García JÁ, Lanceros-Méndez S, Reguera J. Enhancement and Tunability of Plasmonic-Magnetic Hyperthermia through Shape and Size Control of Au:Fe3O4 Janus Nanoparticles. ACS Applied Nano Materials. 2023;6(19):18466-79.
18. Li Z, Aranda-Ramos A, Güell-Grau P, Tajada JL, Pou-Macayo L, Lope Piedrafita S, *et al.* Magnetically amplified photothermal therapies and multimodal imaging with magneto-plasmonic nanodomes. Applied Materials Today. 2018;12:430-40.
19. Su H, Hurd Price CA, Jing L, Tian Q, Liu J, Qian K. Janus particles: design, preparation, and biomedical applications. Mater Today Bio. 2019;4:100033.
20. Duan Y, Zhao X, Sun M, Hao H. Research Advances in the Synthesis, Application, Assembly, and Calculation of Janus Materials. Industrial & Engineering Chemistry Research. 2021;60(3):1071-95.
21. Koya AN, Li L, Li W. Resonant optical trapping of Janus nanoparticles in plasmonic nanoaperture. Applied Physics Letters. 2023;123(22).
22. Ren Y-X, Frueh J, Zhang Z, Rutkowski S, Zhou Y, Mao H, et al. Topologically protected optical pulling force on synthetic particles through photonic nanojet. Nanophotonics. 2024;13(2):239-49.
23. Maccaferri N, Vavassori P, Garoli D. Magnetic control of particle trapping in a hybrid plasmonic nanopore. Applied Physics Letters. 2021;118(19).

24. Ebbesen TW, Lezec HJ, Ghaemi HF, Thio T, Wolff PA. Extraordinary optical transmission through sub-wavelength hole arrays. Nature. 1998;391(6668):667-9.
25. Garcia-Vidal FJ, Martin-Moreno L, Ebbesen TW, Kuipers L. Light passing through subwavelength apertures. Reviews of Modern Physics. 2010;82(1):729-87.
26. Gordon R. Metal Nanoapertures and Single Emitters. Advanced Optical Materials. 2020;8(20).
27. Al Balushi AA, Kotnala A, Wheaton S, Gelfand RM, Rajashekara Y, Gordon R. Label-free free-solution nanoaperture optical tweezers for single molecule protein studies. Analyst. 2015;140(14):4760-78.
28. Koh AL, Fernandez-Dominguez AI, McComb DW, Maier SA, Yang JK. High-resolution mapping of electron-beam-excited plasmon modes in lithographically defined gold nanostructures. Nano Lett. 2011;11(3):1323-30.
29. Rogez B, Marmri Z, Thibaudau F, Baffou G. Thermoplasmonics of metal layers and nanoholes. APL Photonics. 2021;6(10).
30. Yoo D, Gurunatha KL, Choi HK, Mohr DA, Ertsgaard CT, Gordon R, *et al.* Low-Power Optical Trapping of Nanoparticles and Proteins with Resonant Coaxial Nanoaperture Using 10 nm Gap. Nano Lett. 2018;18(6):3637-42.
31. Ahmed R, Butt H. Diffractive Surface Patterns through Single-Shot Nanosecond-Pulsed Laser Ablation. ACS Photonics. 2019;6(7):1572-80.
32. Zhao Y, Hubarevich A, De Fazio AF, Iarossi M, Huang JA, De Angelis F. Plasmonic Bowl-Shaped Nanopore for Raman Detection of Single DNA Molecules in Flow-Through. Nano Lett. 2023;23(11):4830-6.
33. Urban AS, Carretero-Palacios S, Lutich AA, Lohmuller T, Feldmann J, Jackel F. Optical trapping and manipulation of plasmonic nanoparticles: fundamentals, applications, and perspectives. Nanoscale. 2014;6(9):4458-74.
34. Juan ML, Gordon R, Pang Y, Eftekhari F, Quidant R. Self-induced back-action optical trapping of dielectric nanoparticles. Nature Physics. 2009;5(12):915-9.
35. Kudo T, Ishihara H, Masuhara H. Resonance optical trapping of individual dye-doped polystyrene particles with blue- and red-detuned lasers. Opt Express. 2017;25(5):4655-64.
36. Bresolí-Obach R, Kudo T, Louis B, Chang Y-C, Scheblykin IG, Masuhara H, et al. Resonantly Enhanced Optical Trapping of Single Dye-Doped Particles at an Interface. ACS Photonics. 2021;8(6):1832-9.
37. Gordon JP, Ashkin A. Motion of atoms in a radiation trap. Physical Review A. 1980;21(5):1606-17.
38. Huang CH, Kudo T, Bresoli-Obach R, Hofkens J, Sugiyama T, Masuhara H. Surface plasmon resonance effect on laser trapping and swarming of gold nanoparticles at an interface. Opt Express. 2020;28(19):27727-35.
39. Shoji T, Mizumoto Y, Ishihara H, Kitamura N, Takase M, Murakoshi K, et al. Plasmon-Based Optical Trapping of Polymer Nano-Spheres as Explored by Confocal Fluorescence Microspectroscopy: A Possible Mechanism of a Resonant Excitation Effect. Japanese Journal of Applied Physics. 2012;51(9R).
40. Shoji T, Tsuboi Y. Plasmonic Optical Tweezers toward Molecular Manipulation: Tailoring Plasmonic Nanostructure, Light Source, and Resonant Trapping. J Phys Chem Lett. 2014;5(17):2957-67.

41. Jiang Q, Rogez B, Claude J-B, Baffou G, Wenger J. Temperature Measurement in Plasmonic Nanoapertures Used for Optical Trapping. ACS Photonics. 2019;6(7):1763-73.
42. Praveen Kamath P, Sil S, Truong VG, Nic Chormaic S. Particle trapping with optical nanofibers: a review [Invited]. Biomed Opt Express. 2023;14(12):6172-89.
43. Kotnala A, Zheng Y. Opto-thermophoretic fiber tweezers. Nanophotonics. 2019;8(3):475-85.
44. Xin H, Cheng C, Li B. Trapping and delivery of Escherichia coli in a microfluidic channel using an optical nanofiber. Nanoscale. 2013;5(15):6720-4.
45. Erb RM, Jenness NJ, Clark RL, Yellen BB. Towards holonomic control of Janus particles in optomagnetic traps. Adv Mater. 2009;21(47):4825-9.
46. Brennan G, Bergamino S, Pescio M, Tofail SAM, Silien C. The Effects of a Varied Gold Shell Thickness on Iron Oxide Nanoparticle Cores in Magnetic Manipulation, T(1) and T(2) MRI Contrasting, and Magnetic Hyperthermia. Nanomaterials (Basel). 2020;10(12).
47. Zhang L, Luo Q, Zhang F, Zhang D-M, Wang Y-S, Sun Y-L, et al. High-performance magnetic antimicrobial Janus nanorods decorated with Ag nanoparticles. Journal of Materials Chemistry. 2012;22(45).
48. Zhang H, Zhu T, Li M. Quantitative Analysis of the Shape Effect of Thermoplasmonics in Gold Nanostructures. J Phys Chem Lett. 2023;14(16):3853-60.
49. Cunha J, Guo TL, Koya AN, Toma A, Prato M, Della Valle G, et al. Photoinduced Temperature Gradients in Sub‐Wavelength Plasmonic Structures: The Thermoplasmonics of Nanocones. Advanced Optical Materials. 2020;8(18).
50. Cunha J, Guo TL, Della Valle G, Koya AN, Proietti Zaccaria R, Alabastri A. Controlling Light, Heat, and Vibrations in Plasmonics and Phononics. Advanced Optical Materials. 2020;8(24).
51. Chen J, Ye Z, Yang F, Yin Y. Plasmonic Nanostructures for Photothermal Conversion. Small Science. 2021;1(2).
52. Koya AN, Cunha J, Guerrero‐Becerra KA, Garoli D, Wang T, Juodkazis S, et al. Plasmomechanical Systems: Principles and Applications. Advanced Functional Materials. 2021;31(41).
53. Huergo MA, Schuknecht F, Zhang J, Lohmüller T. Plasmonic Nanoagents in Biophysics and Biomedicine. Advanced Optical Materials. 2022;10(14).
54. Xuan M, Wu Z, Shao J, Dai L, Si T, He Q. Near Infrared Light-Powered Janus Mesoporous Silica Nanoparticle Motors. J Am Chem Soc. 2016;138(20):6492-7.
55. Peng X, Chen Z, Kollipara PS, Liu Y, Fang J, Lin L, et al. Opto-thermoelectric microswimmers. Light Sci Appl. 2020;9:141.
56. Ruhoff VT, Arastoo MR, Moreno-Pescador G, Bendix PM. Biological Applications of Thermoplasmonics. Nano Lett. 2024;24(3):777-89.
57. Cui X, Ruan Q, Zhuo X, Xia X, Hu J, Fu R, et al. Photothermal Nanomaterials: A Powerful Light-to-Heat Converter. Chem Rev. 2023;123(11):6891-952.
58. Sipova-Jungova H, Andren D, Jones S, Kall M. Nanoscale Inorganic Motors Driven by Light: Principles, Realizations, and Opportunities. Chem Rev. 2020;120(1):269-87.
59. Ilic O, Kaminer I, Zhen B, Miller OD, Buljan H, Soljačić M. Topologically enabled optical nanomotors. Science Advances. 2017;3(6):e1602738.

60. Ding H, Kollipara PS, Kim Y, Kotnala A, Li J, Chen Z, et al. Universal optothermal micro/nanoscale rotors. Science Advances. 2022;8(24):eabn8498.
61. Ilic O, Kaminer I, Lahini Y, Buljan H, Soljačić M. Exploiting Optical Asymmetry for Controlled Guiding of Particles with Light. ACS Photonics. 2016;3(2):197-202.
62. Koya AN, Cunha J, Guo TL, Toma A, Garoli D, Wang T, et al. Novel Plasmonic Nanocavities for Optical Trapping ‐ Assisted Biosensing Applications. Advanced Optical Materials. 2020;8(7).
63. Rodríguez-Sevilla P, Arita Y, Liu X, Jaque D, Dholakia K. The Temperature of an Optically Trapped, Rotating Microparticle. ACS Photonics. 2018;5(9):3772-8.
64. Karpinski P, Jones S, Sipova-Jungova H, Verre R, Kall M. Optical Rotation and Thermometry of Laser Tweezed Silicon Nanorods. Nano Lett. 2020;20(9):6494-501.
65. Farzin A, Etesami SA, Quint J, Memic A, Tamayol A. Magnetic Nanoparticles in Cancer Therapy and Diagnosis. Adv Healthc Mater. 2020;9(9):e1901058.
66. Espinosa A, Reguera J, Curcio A, Munoz-Noval A, Kuttner C, Van de Walle A, et al. Janus Magnetic-Plasmonic Nanoparticles for Magnetically Guided and Thermally Activated Cancer Therapy. Small. 2020;16(11):e1904960.
67. Huang J, Guo M, Ke H, Zong C, Ren B, Liu G, et al. Rational Design and Synthesis of gamma Fe2O3@Au Magnetic Gold Nanoflowers for Efficient Cancer Theranostics. Adv Mater. 2015;27(34):5049-56.
68. Hu XQ, Chen JR, Xiao WJ. Controllable Remote C-H Bond Functionalization by Visible-Light Photocatalysis. Angew Chem Int Ed Engl. 2017;56(8):1960-2.
69. Chicheł A, Skowronek J, Kubaszewska M, Kanikowski M. Hyperthermia – description of a method and a review of clinical applications. Reports of Practical Oncology & Radiotherapy. 2007;12(5):267-75.
70. Bienia A, Wiechec-Cudak O, Murzyn AA, Krzykawska-Serda M. Photodynamic Therapy and Hyperthermia in Combination Treatment-Neglected Forces in the Fight against Cancer. Pharmaceutics. 2021;13(8).
71. Beik J, Abed Z, Ghoreishi FS, Hosseini-Nami S, Mehrzadi S, Shakeri-Zadeh A, et al. Nanotechnology in hyperthermia cancer therapy: From fundamental principles to advanced applications. J Control Release. 2016;235:205-21.
72. Fatima H, Charinpanitkul T, Kim KS. Fundamentals to Apply Magnetic Nanoparticles for Hyperthermia Therapy. Nanomaterials (Basel). 2021;11(5).
73. Jose J, Kumar R, Harilal S, Mathew GE, Parambi DGT, Prabhu A, et al. Magnetic nanoparticles for hyperthermia in cancer treatment: an emerging tool. Environ Sci Pollut Res Int. 2020;27(16):19214-25.
74. Zuo S, Wang J, An X, Zhang Y. Janus Magnetic Nanoplatform for Magnetically Targeted and Protein/Hyperthermia Combination Therapies of Breast Cancer. Front Bioeng Biotechnol. 2021;9:763486.
75. Ali I, Chen J, Ahmed Khan S, Jamil Y, Shah AA, Shah AK, et al. Photothermal Hyperthermia Study of Ag/Ni and Ag/Fe Plasmonic Particles Synthesized Using Dual-Pulsed Laser. Magnetochemistry. 2023;9(3).
76. Wang Z, Zhang F, Shao D, Chang Z, Wang L, Hu H, et al. Janus Nanobullets Combine Photodynamic Therapy and Magnetic Hyperthermia to Potentiate Synergetic Anti-Metastatic Immunotherapy. Adv Sci (Weinh). 2019;6(22):1901690.

77. Wu Q, Lin Y, Wo F, Yuan Y, Ouyang Q, Song J, et al. Novel Magnetic-Luminescent Janus Nanoparticles for Cell Labeling and Tumor Photothermal Therapy. Small. 2017;13(39).

78. Fiorito S, Soni N, Silvestri N, Brescia R, Gavilan H, Conteh JS, *et al.* Fe(3)O(4) @Au@Cu(2-)(x) S Heterostructures Designed for Tri-Modal Therapy: Photo- Magnetic Hyperthermia and (64) Cu Radio-Insertion. Small. 2022;18(18):e2200174.

79. Karatum O, Han M, Erdogan ET, Karamursel S, Nizamoglu S. Physical mechanisms of emerging neuromodulation modalities. J Neural Eng. 2023;20(3).

80. Han M, Srivastava SB, Yildiz E, Melikov R, Surme S, Dogru-Yuksel IB, *et al.* Organic Photovoltaic Pseudocapacitors for Neurostimulation. ACS Appl Mater Interfaces. 2020;12(38):42997-3008.

81. Han M, Karatum O, Nizamoglu S. Optoelectronic Neural Interfaces Based on Quantum Dots. ACS Appl Mater Interfaces. 2022;14(18):20468-90.

82. Han M, Yildiz E, Bozuyuk U, Aydin A, Yu Y, Bhargava A, et al. Janus microparticles-based targeted and spatially-controlled piezoelectric neural stimulation via low-intensity focused ultrasound. Nat Commun. 2024;15(1):2013.

83. Mousavi NSS, Ramadi KB, Song Y-A, Kumar S. Plasmonics for neuroengineering. Communications Materials. 2023;4(1).

84. Gao Y, Guo Y, Yang Y, Tang Y, Wang B, Yan Q, et al. Magnetically Manipulated Optoelectronic Hybrid Microrobots for Optically Targeted Non-Genetic Neuromodulation. Adv Mater. 2024;36(8):e2305632.

85. Liu Y, Yi Z, Yao Y, Guo B, Liu X. Noninvasive Manipulation of Ion Channels for Neuromodulation and Theranostics. Accounts of Materials Research. 2022;3(2):247-58.

86. Chen L, Mo C, Wang L, Cui H. Direct numerical simulation of the self-propelled Janus particle: use of grid-refined fluctuating lattice Boltzmann method. Microfluidics and Nanofluidics. 2019;23(5).

87. Kohl R, Corona E, Cheruvu V, Veerapaneni S. Fast and accurate solvers for simulating Janus particle suspensions in Stokes flow. Advances in Computational Mathematics. 2023;49(4).

88. Koya A N, Zhu X, Ohannesian N, Yanik A A, Alabastri A, Proietti Zaccaria R, Krahne R, Shih W-C, Garoli D. Nanoporous Metals: From Plasmonic Properties to Applications in Enhanced Spectroscopy and Photocatalysis. ACS Nano 2021, 15, 4, 6038–6060.

89. Lehmuskero A, Johansson P, Rubinsztein-Dunlop H, Tong L, Käll M. Laser trapping of colloidal metal nanoparticles. ACS Nano 2015, 9, 4, 3453–3469.

90. Ashkin A, Dziedzic JM, Smith PW. Continuous-wave self-focusing and self-trapping of light in artificial Kerr media. Optics Letters 1982, 7, 6, 276–278.

91. Ashkin A. Acceleration and trapping of particles by radiation pressure. Physical Review Letters 1970, 24, 4, 156.

92. Park S, Choi J, Ko N, Mondal S, Pal U, Lee BI, Oh J. Beta cyclodextrin conjugated $AuFe_3O_4$ Janus nanoparticles with enhanced chemo-photothermal therapy performance. Acta Biomaterialia 2024, 182, 213–227.

93. Shi XH, Tao L, Wang L, Liu X, Liu SL, Wang ZG. Plasmonic-fluorescent Janus Au-PbS nanoparticles with bright near-infrared-IIb fluorescence and photothermal effect for computed tomography imaging-guided combination cancer therapy. Chemistry of Materials 2024, 36, 6, 2776–2789.

94. Zhang X, Wang W, Su L, Ge X, Ye J, Zhao C, He Y, Yang H, Song J, Duan H. Plasmonic-fluorescent Janus $Ag/Ag_2S$ nanoparticles for in situ $H_2O_2$-activated NIR-II fluorescence imaging. Nano Letters 2021, 21, 6, 2625–2633.

95. Bao J, Liu R, Yu Z, Cheng Z, Chang B. Activatable Janus nanoparticles for precise NIR-II bioimaging and synergistic cancer therapy. Advanced Functional Materials 2024, 34, 27, 2316646.

96. Zhang X, Wang W, Su L, Ge X, Ye J, Zhao C, He Y, Yang H, Song J, Duan H. Plasmonic-fluorescent Janus $Ag/Ag_2S$ nanoparticles for in situ $H_2O_2$-activated NIR-II fluorescence imaging. Nano Letters 2021, 21, 6, 2625–2633.

97. Reguera J, Jiménez de Aberasturi D, Winckelmans N, Langer J, Bals S, Liz-Marzán LM. Synthesis of Janus plasmonic–magnetic, star–sphere nanoparticles, and their application in SERS detection. Faraday Discussions 2016, 191, 47–59.

98. Malmir K, Okell W, Trichet AP, Smith MJ. Characterization of nanoparticle size distributions using a microfluidic device with integrated optical microcavities. Lab Chip, 2022, 22, 3499–3507.